# Intricacies of Strain and Magnetic Field Induced Charge Order Melting in $Pr_{0.5}Ca_{0.5}MnO_3$ Thin Films

*Dipak Kumar Baisnab[*], T. Geetha Kumary, A. T. Satya, Awadhesh Mani, J. Janaki, R. Nithya, L. S. Vaidhyanathan, M. P. Janawadkar and A. Bharathi*

Materials Science Group,
Indira Gandhi Centre for Atomic Research,
Kalpakkam 603102, India

**Abstract**

Thin films of the half doped manganite $Pr_{0.5}Ca_{0.5}MnO_3$ were grown on (100) oriented MgO substrates by pulsed laser deposition technique. In order to study the effect of strain on the magnetic field induced charge order melting, films of different thicknesses were prepared and their properties were studied by x-ray diffraction, electrical resistivity and magnetoresistance measurements. A field induced charge order melting is observed for films with very small thicknesses. The charge order transition temperature and the magnetic filed induced charge order melting are observed to depend on the nature of strain that is experienced by the film.

---

[*] corresponding author, email: dkbb@igcar.gov.in

# INTRODUCTION

The rare earth doped manganites $R_{1-x}A_xMnO_3$ (R = trivalent rare earth ion, A = divalent alkaline earth ion) have been studied extensively[1-3] because of their exciting physical properties arising due to the simultaneous interplay of spin, charge and orbital degrees of freedom. The studies on manganites still continue to attract attention not only due to the richness in their physical properties, but also due to their potential in technological applications[3]. The Colossal Magnetoresistance (CMR) effect is one of the spectacular properties of these materials which have a huge potential for a variety of applications ranging from read head technology to contactless potentiometers and bolometers.[3] However, the realization of these applications is more viable in thin film forms, because the physical property suitable for a particular application can be tailored in a controlled fashion in the thin film form by the use of appropriate external parameters[4-7] such as magnetic field and/or substrate generated strain, irradiation by a high flux of photons, external pressure etc. In recent years, strain engineering has emerged as a promising technique for inducing charge order melting in thin films of half doped ($R_{0.5}A_{0.5}MnO_3$) manganites. The charge order (CO) state is a long range ordering of $Mn^{3+}$ and $Mn^{4+}$ ions. It is known that in perovskite manganese oxides ($R_{1-x}A_xMnO_3$), the double exchange interaction is mediated by the overlap of the 2p orbitals of O and the $e_g$ orbitals of Mn ions. This structure is highly influenced by the tolerance factor $\Gamma = (r_A+r_O)/\sqrt{2}\,(r_{Mn}+r_O)$, where $r_A$, $r_O$ and $r_{Mn}$ denote the radii of the respective ions. The 180º bond angle value of Mn-O-Mn gets perturbed upon decreasing $\Gamma$, which can cause narrowing of the effective one-electron bandwith W of the $e_g$ band[7]. This reduction of W leads to destabilization of the double exchange mediated ferromagnetic (FM) state due to the presence of competing instabilities such as charge-orbital ordering (CO-OO), antiferromagnetic superexchange interaction etc. In manganite system, two competing phases, viz., FM and CO-OO, can form across the first order phase transition boundary in the neighbourhood of the bicritical point. In this case, one can expect a sudden change from one phase to the other by controlled perturbation.

These perturbations can be in the form of doping by particular atoms to alter the lattice parameter or by introducing strain to regulate the lattice distortion. Lattice mismatch generated strain can transform the CO-OO phase to conducting FM phase. This is known as melting of charge order by introducing disorder into the CO state.

$Pr_{0.5}Ca_{0.5}MnO_3$ (PCMO) is a paramagnetic insulator at high temperature and undergoes a phase transition into a charge ordered insulating state below around 240 K. A CE type antiferromagnetic order sets in at still lower temperatures (viz., ~170K) [8]. As discussed earlier it is possible to destroy the charge ordered state and induce an insulator to metal transition[10] by the application of an external magnetic field. The destruction of the CO state is accompanied by a magnetic transition and a CMR effect.[11] A magnetic field of around 25 T is required to melt the CO state in bulk PCMO[11]. On the other hand CO melting occurs at a lower magnetic field[12-15] in the case of thin films. The effect of tensile and compressive strains induced by the substrate lattice mismatch on the CO melting in PCMO has already been studied. In PCMO films with tensile strains, formed on $SrTiO_3$ (STO) substrates, CO melting occurs at much lower fields (depending on the thickness of the film) compared to the bulk, whereas in the films with compressive strains grown on $LaAlO_3$ (LAO) substrates, magnetic field is reported to have a weaker effect[9, 16-18]. In the present paper, we report on the observation of charge order melting for very thin films of PCMO grown on highly lattice mismatched MgO substrate. The lattice mismatch $\delta$ defined by $\delta = (a_s - a_T)/a_s$, (where $a_s$ and $a_T$ are the lattice parameters of the substrate and bulk PCMO) is calculated to be + 9.4 % for MgO, which corresponds to a large tensile strain. This is very large compared to the value of $\delta$ for STO (+ 2.25 %: tensile strain) or LAO (– 0.77 %: compressive strain) substrates.[3] So we expect a large effect of strain on the CO state of PCMO in the films deposited on the MgO substrate. Our studies show that the orientation of the film also dictates the nature of strain and hence the CO melting.

**EXPERIMENTAL DETAILS**

The bulk polycrystalline target material of $Pr_{0.5}Ca_{0.5}MnO_3$ used for the preparation of thin films was synthesized by the standard solid state reaction method. Stoichiometric quantities of $Pr_6O_{11}$, $CaCO_3$ and $MnO_2$ were heat treated several times in the temperature range 1200 – 1350 ºC to obtain the phase pure compound. Thin films, with nominal thickness ranging from 25 – 200 nm, were grown by the pulsed laser deposition (PLD) technique on (100) oriented MgO substrates. This substrate was chosen because of its higher lattice mismatch with that of PCMO. KrF excimer laser (wave length: 248 nm) with a fluence of ~ 1.1 $J/cm^2$ and pulse repetition rate of 5 Hz was used for the preparation of all the films. Laser deposition was carried out under an oxygen pressure of 0.15 mbar with a flow rate of 40 sccm, after obtaining a base pressure of ~ $3 \times 10^{-5}$ mbar in the chamber. The substrate temperature was maintained at 800 ºC and the substrate to target distance was kept at 4 cm during the deposition. After the deposition, the thin films were allowed to cool to room temperature in ~ 2 hours under 1 bar oxygen pressure.

The thickness and surface roughness of the films were measured using DEKTAK surface profiler. The crystal structure and the orientation of the films and the bulk polycrystalline sample were characterized by X-ray diffraction (XRD) measurements in a STOE diffractometer using Cu $K_\alpha$ radiation. Electrical resistivity measurements were carried out in the temperature range 4.2 – 300 K using the four probe technique in a dipstick cryostat. Magnetoresistance (MR) measurements for all the thin film samples were carried out with the magnetic field (up to 12 T) applied parallel to the plane of the film. MR of 35 nm film was measured by applying the field parallel as well as perpendicular to the plane of the film in order to check for any anisotropy in magnetoresistance. A calibrated cernox resistance thermometer was used for the temperature measurements in the presence of magnetic field.

**RESULTS AND DISCUSSION**

Figure 1 shows the XRD pattern of bulk PCMO. All the lines could be indexed to an orthorhombic unit cell with lattice parameters, $a$ = 5.3912 Å, $b$ = 7.6304 Å and $c$ = 5.4011 Å. Magnetisation was measured down to a temperature of 10 K using a liquid helium based vibrating sample magnetometer (VSM) operating at 20.4 Hz. The characteristic signals associated with the CO transition at $T_{CO}$ = 240 K and the AFM transition at $T_N$ = 150 K are clearly seen (data shown in the inset of Fig. 1).

The thicknesses of the thin films were measured to be 25, 35, 42, 65, 70, 120 and 200 nm using the DEKTAK surface profiler (designated henceforth as P25, P35,……., P200 respectively, according to the thickness of the films). The average surface roughness of the films was found to vary from 5.5 nm to 6.5 nm. XRD measurements indicated that the PCMO films had a (121) orientation (*i.e.*, [121] axis perpendicular to the plane of the substrate). Figure 2 represents the XRD patterns of a few representative thin films. Very small intensity lines indexed as (202) and (240) are also observed in some thin films as shown in Fig.2.

The temperature dependent resistivity ($\rho(T)$) for a few films along with the bulk polycrystalline PCMO is shown in Fig. 3. The resistivity shows a semiconducting like behavior in the entire temperature range studied. A systematic reduction in resistivity is observed with decrease in film thickness indicating that the changes in the resistivity are intrinsic to the film. A variable range hopping type of conductivity (Resistivity, $\rho \propto exp\ (T_0/T)^{0.25}$ where $T_0$ is hopping parameter) is observed at low temperatures for the bulk and thin film samples as shown in the inset of Fig. 3. The $\rho$ vs $1/T^{0.25}$ plots for the bulk, and the thin films P120 and P35, drawn on a semilogarithmic scale, are shown in the inset of fig.3. For the bulk sample, there is a deviation from the linear behavior near ~ 240 K, which is identified as the $T_{CO}$ above which a thermally activated behavior for conductivity ($\rho \propto exp\ (\Delta E/k_BT)$ where $\Delta E$ is energy gap and $k_B$ is *the* Boltzmann constant ) is observed. In the samples

P120 and P35, the slope change is observed at around ~ 225 K and ~ 195 K respectively, as shown in the inset of Fig. 3. We could not obtain the $T_{CO}$ of thin films from the magnetization measurements as the signal levels were below the detection limit of the VSM. Yang *et. al*[19] have estimated the $T_{CO}$ of thin films by considering the slope changes in the resistivity plot. Following this report, if we assume the slope change in $\rho$ vs $1/T^{0.25}$ to be due to the CO transition, $T_{CO}$ seems to decrease with decrease in the film thickness. Even though it is difficult to obtain the exact temperature at which the slope change occurs, the decrease in $T_{CO}$ with decrease in film thickness is quite apparent from the plot shown in the inset of Fig.3. If that is the case, the systematic decrease in resistivity observed with decrease in film thickness can be attributed to the stabilization of CO at a lower temperature as compared to that in the bulk sample.

Magnetoresistance (MR) measurements were carried out for all the films by applying magnetic fields up to 12 T. Films with 35 and 42 nm thicknesses showed an insulator to metal transition (MIT) on application of magnetic field. On the other hand, films with higher thicknesses and the one with the lowest thickness (P25) did not show the MIT in the range of temperature and magnetic fields investigated. A negative MR is observed in all these films at low temperatures. Figure 4 (a) represents the temperature dependent resistance of P35 with the magnetic field applied parallel to the plane of the film (*i.e.*, parallel to the (121) plane). An insulator to metal transition, corresponding to the CO melting, occurs under magnetic field. As can be seen from Fig. 4 (a) a metallic behavior in resistivity is observed in P35 even under zero field at temperatures below ~ 50 K. The resistance data for the cooling and warming cycles are marked by the arrows in the figure. A small hysteresis, present between the cooling and warming cycles, decreases with increase in the magnetic field. This result is at variance with the reported observations of other workers who have investigated the bulk and thin films of PCMO deposited on STO substrate.[9] The temperature corresponding to the MIT increases with increase in magnetic field. Magnetoresistance measurements for P35 were also carried out by applying the field perpendicular to (121) plane, the results of which are shown in Fig.4(b). The absence of a significant anisotropy is evident from a comparison of Fig.4(a) and (b). Fig.4(c) depicts the MR behaviour of the P42 film. Under zero field, MIT is not observed for P42 unlike the behaviour exhibited by P35. However, under the application of magnetic fields, P42 also exhibits the MIT as shown in Fig 4 (c). P42 also shows a small hysteresis between the cooling and warming cycles, which decreases with increase in the applied field.

MR is also measured at fixed temperatures by changing the magnetic field from 0 - 12 T. The result of the field scan carried out at 52 K is shown in Fig. 5 (a) and 5 (b) for P35 and P42 respectively. Both the films are insulating and exhibit very high values of resistivity at 52 K under zero field. As shown in Fig.5, conductivity becomes measurable in our set-up above a field of ~ 6 T for P35 and ~ 8 T for P42 and the resistance decreases for both the films with increase in magnetic field. Interestingly, the conductivity remains measurable in P42 till 6 T when the field is reduced from 12 T due to the hysteresis.

The films with higher thicknesses did not show the MIT even under the application of a field of 12 T. The resistance data for P65 and P200 taken under different magnetic fields are shown in Fig. 6 (a) and (b) respectively. A negative MR is clearly observed for non-zero external magnetic field. The value of MR is observed to decrease with increase in the film thickness. As mentioned earlier, the film with lowest thickness, P25, did not show the MIT under the application of magnetic field. It may be noted that the MIT has not been observed by other workers in very thin films[8, 12] of PCMO grown on STO substrates. On the other hand thicker films show a field induced MIT and it has been observed that the MIT occurs for smaller fields when the thickness of the film is increased. The unusual behaviour of the thin films in the present study appears to be on account of the fact that it is not the strain as such that is responsible for the CO melting in thin films but it is the strain relaxation, that leads to atomic scale disorder in the system, giving rise to the CO melting. So the very thin 'strained' films do not show CO melting whereas the 'strain relaxed' thicker films undergo CO melting under relatively small applied fields. The effect of strain relaxation on the CO melting has also been studied[12] by post-annealing the films, and a decrease in the CO melting field with increase in the annealing time has been reported. Also, an enhancement in CO melting temperature[19] has been reported in strained thin films, grown on STO, which suggests stabilization of CO with a tensile strain. Since MgO has a larger lattice mismatch (9.4%) with PCMO[3] compared to STO (+ 2.25 %), one can expect to observe a large effect on the CO melting field for films grown on MgO. But we observed the CO melting only for films in a narrow thickness range and the CO melting field was observed to increase with increase in thickness, contrary to what is reported for films grown on STO. According to our XRD results, the PCMO films have grown with a (121) orientation whereas those grown on STO are reported to have (010) orientation.[8, 12] Hence one of the reasons for the difference between our results and those in the literature could be that the large strain that is calculated considering the $a$ and $c$ lattice parameters of PCMO is probably not present in our films due to the difference in the orientation of the film. There is a report of field induced MIT under the application of a magnetic field of 5 T for PCMO film with 200 nm thickness grown on MgO substrate.[13] But these films have (001) and (110) orientations as seen by the XRD. Hence, the orientation of the films may be playing a crucial role in controlling the strain and hence the CO melting. It is possible to have a small compressive strain, instead of the tensile strain as expected, in thin films formed with the (121) orientation. According to Yang et. al,[19] tensile strain leads (in PCMO films grown on STO) to an increase in $T_{CO}$ whereas a compressive strain (in PCMO films grown on LAO) causes a decrease in $T_{CO}$. We have observed a decrease in $T_{CO}$ with decrease in film thickness, for the films grown on MgO, from the analysis of the resistivity data. Our Resistivity and magnetoresistance data suggest a compressive strain in the films grown with (121) orientation. Additional experiments are required to understand the role of the microstructure, the exact nature of the strain and the relation between the effect of strain and CO melting in these films.

## CONCLUSIONS

$Pr_{0.5}Ca_{0.5}MnO_3$ films with thickness ranging form 25 to 200 nm were prepared by pulsed laser deposition technique. XRD measurements show (121) orientation for all the films studied. A decrease in resistivity and the charge order transition temperature is observed with decrease in the film thickness. A field induced insulator to metal transition corresponding to the charge order melting is observed for films with very small thickness (P35, P42). However, it appears that there is a range of film thickness (35 nm to 42nm in present study) for which CO melting is observed; films with thickness beyond this range do not exhibit MIT under applied magnetic fields upto 12 T. The film with 35 nm thickness exhibits a MIT even under zero field at low temperatures. It is suggested that the charge order melting and the observed variation in $T_{CO}$ could understood based on the strain relaxation effects arising from the [121] orientation of the PCMO films deposited on MgO substrates.

## ACKNOWLEDGMENT

The authors thank Mr. Shilpam Sharma for providing the magnetization data of the bulk PCMO and Dr C.S. Sundar for constant support and encouragement during the course of this work.

**Figure Captions**

Fig. 1 The XRD pattern of the polycrystalline bulk PCMO. The temperature dependent magnetization data is shown in the inset. $T_{CO}$ and $T_N$ are marked using arrows.

Fig. 2 The XRD patterns of the films with thicknesses (a) 200 nm, (b) 65 nm and (c) 35 nm. All the films have (121) orientation.

Fig. 3 Resistivity *vs* temperature data for the bulk as well as a few representative thin films. The semilogarithmic plots of resistivity as function of $1/T^{0.25}$ for the bulk as well as the films P120 and P35 are shown in the inset.

Fig. 4 Temperature dependent resistance data under different magnetic fields for the films P35 with field (a) parallel and (b) perpendicular to the (121) plane and for (c) P42 with field parallel to the (121) plane. The cooling and warming cycles are marked with arrows.

Fig. 5 The resistance measured by scanning the magnetic field from 0-12 T at 52K for (a) P35 and (b) P42 films. The data corresponding to the increasing and decreasing field scans for P42 are shown indicating a clear hysteresis.

Fig. 6 The temperature dependent resistance data for (a) P65 and (b) P200. A negative MR is observed for both the films.

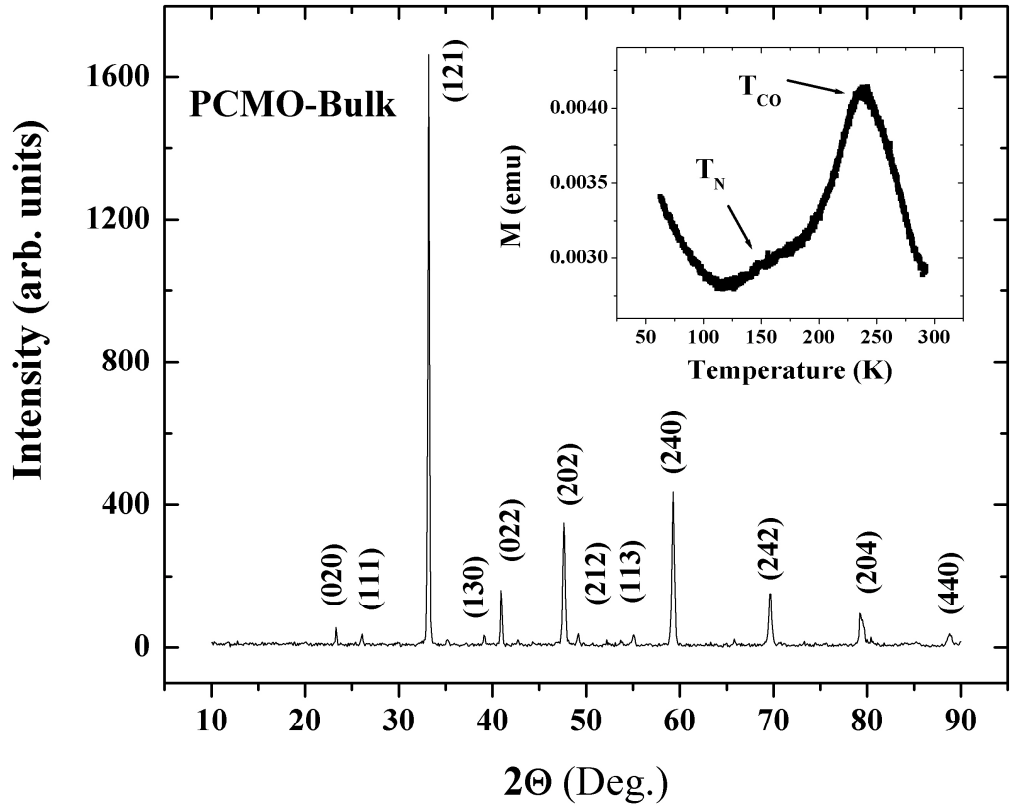

Fig. 1

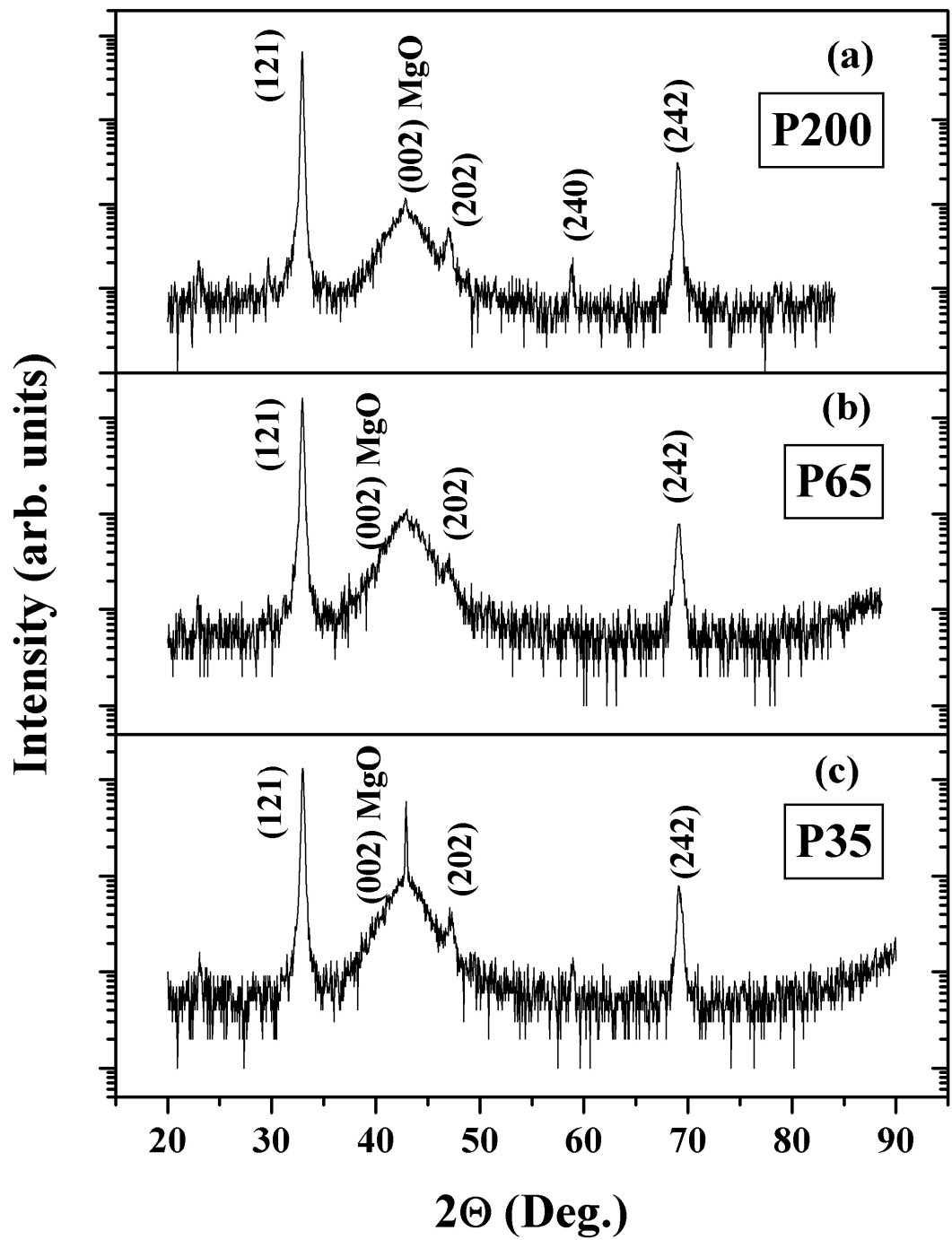

Fig. 2

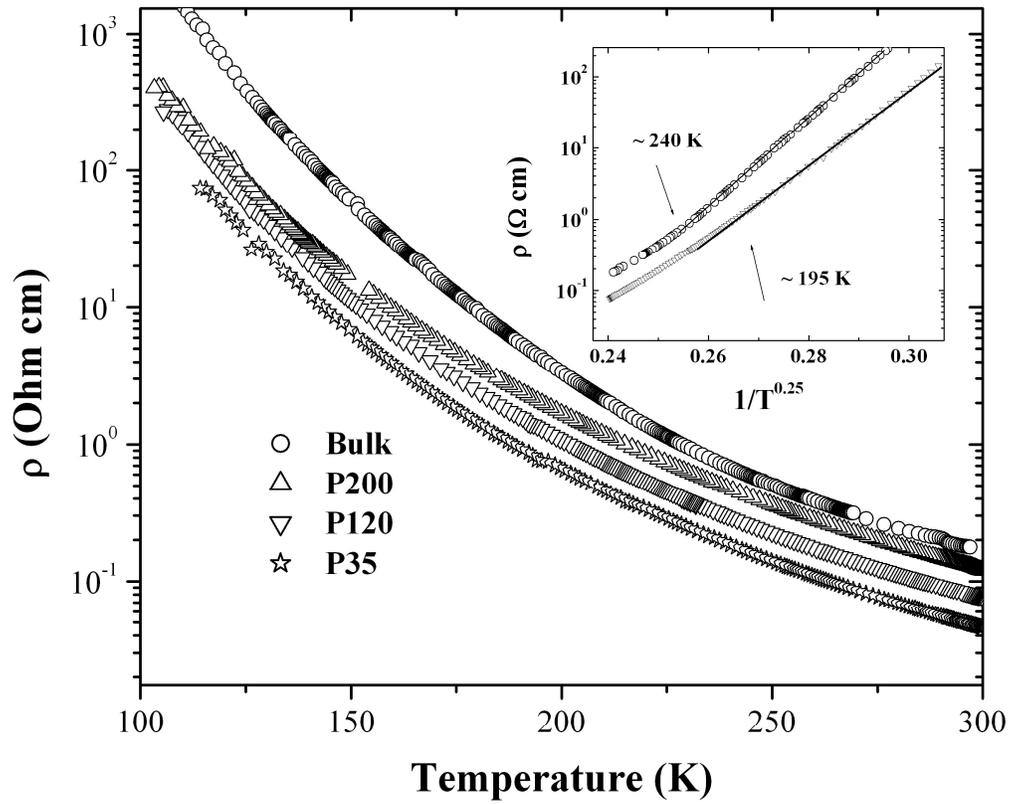

Fig. 3

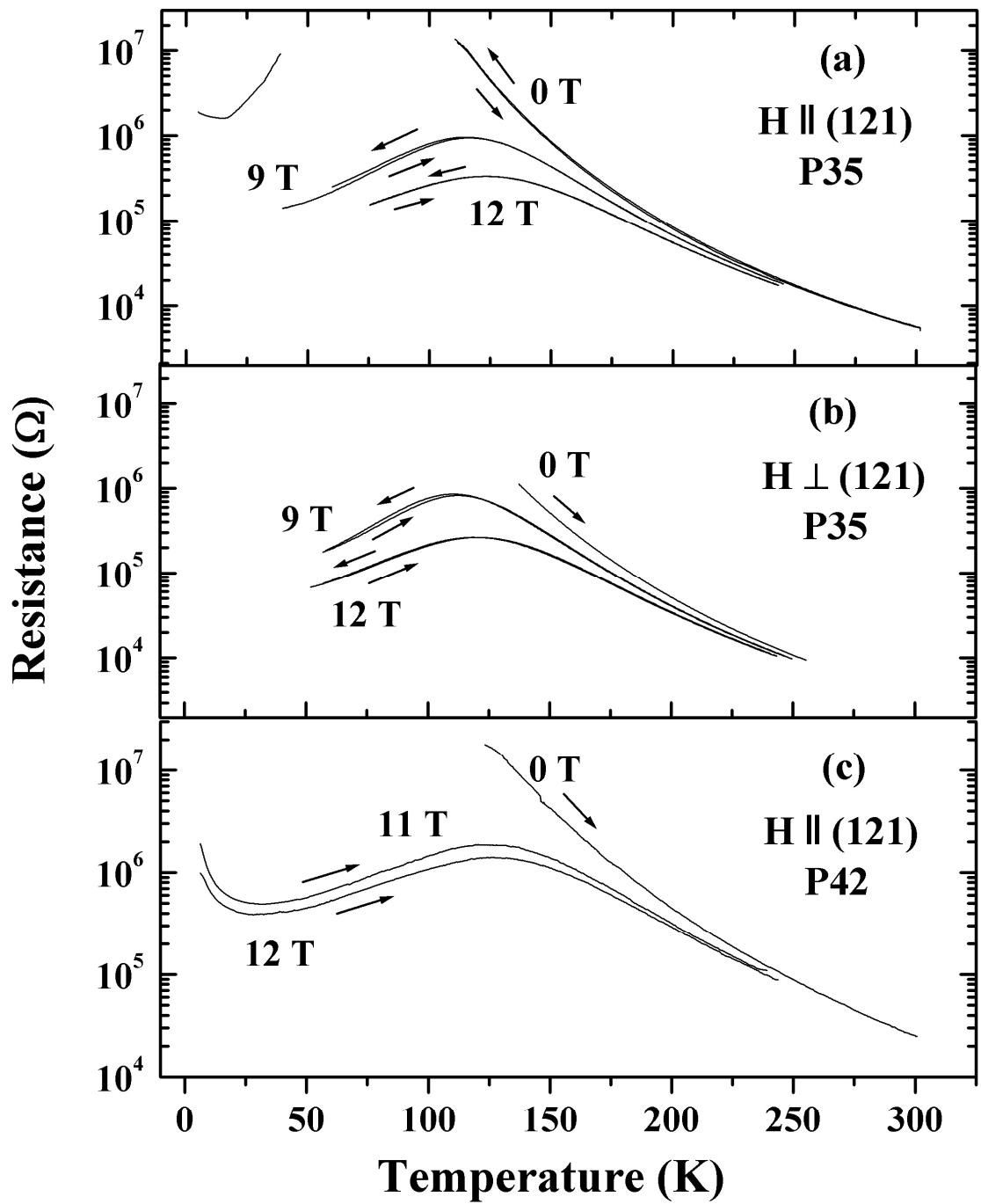

Fig. 4

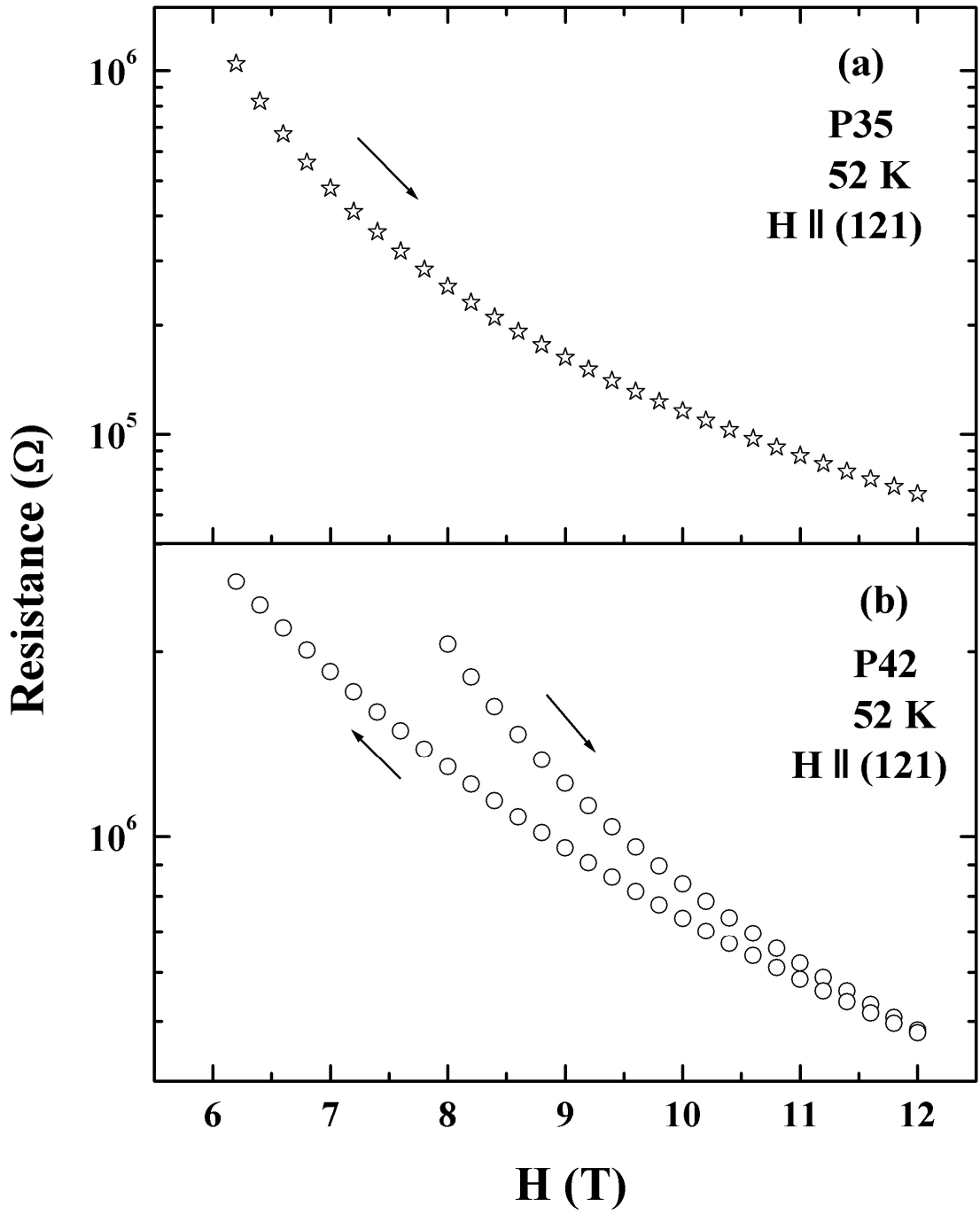

Fig. 5

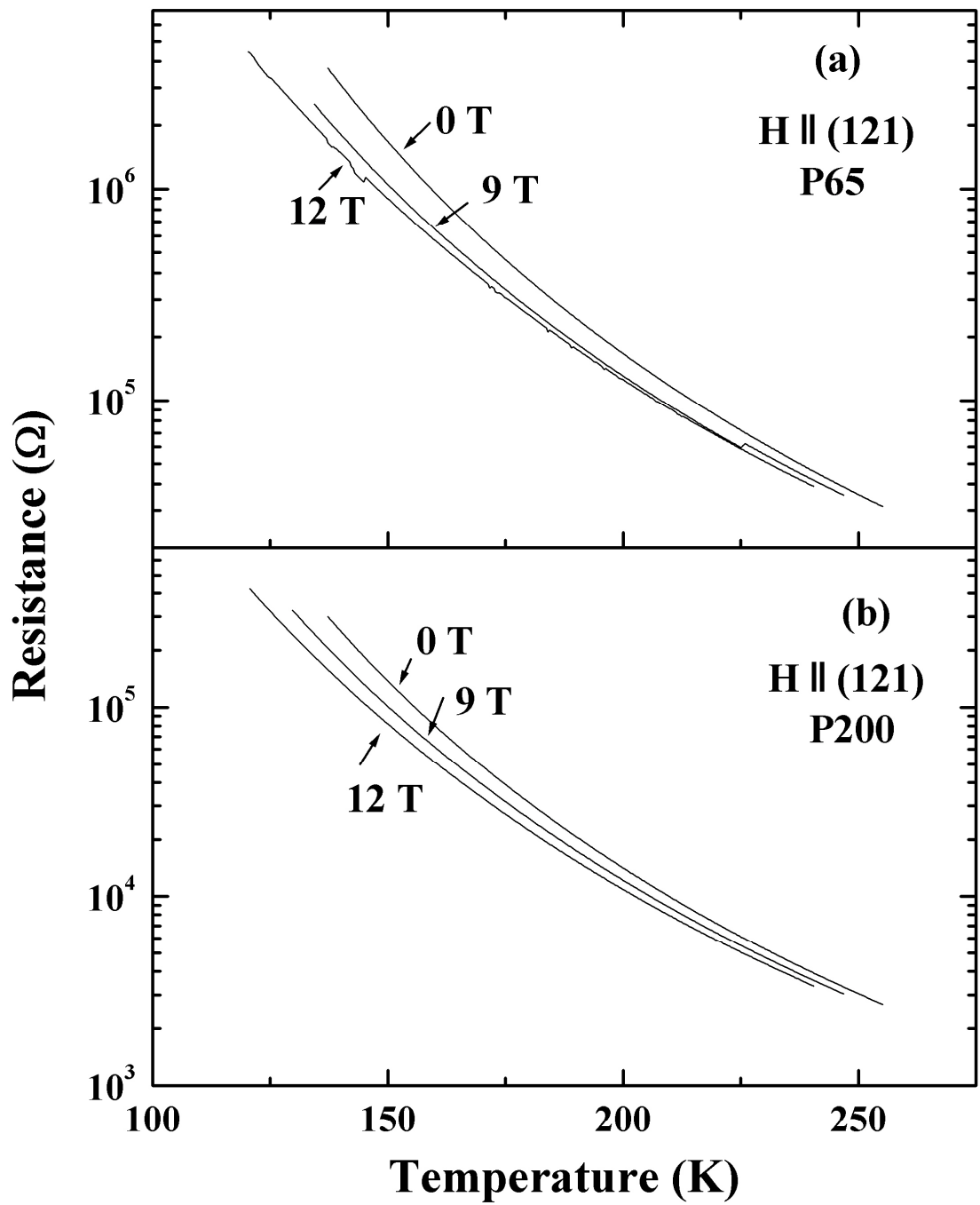

Fig. 6